\documentclass[12pt]{iopart}

\usepackage{graphicx}
\begin{document}

\title{Yields and elliptic flow of $d(\overline{d})$ and $^{3}He(\overline{^{3}He})$ in
 Au+Au collisions at $\sqrt{s_{_{NN}}} = 200\ GeV$}

\author{Haidong Liu\footnote{Supported by National Natural Science Foundation of China (10475074 10620120286)} (for the STAR Collaboration)}

\address{Department of Modern Physics,
University of Science \& Technology of China, Hefei, 230026, P. R.
China}
\ead{hdliu@mail.ustc.edu.cn}
\begin{abstract}
We present the transverse momentum spectra at mid-rapidity for $d$,
$\overline{d}$ ($1<p_{T}<4\ GeV/c$) and $^{3}He$,
$\overline{^{3}He}$ ($2<p_{T}<6\ GeV/c$) measured by the STAR
experiment at RHIC and extract the coalescence parameters $B_{2}$
and $B_{3}$ (respectively). We also present the $v_{2}$ measurement
for $d(\overline{d})$ and $^{3}He(\overline{^{3}He})$. We find that
the $d(\overline{d})$ and $p(\overline{p})$ $v_{2}$ follows the
atomic mass number A scaling within errors and a negative $v_{2}$
has been observed for $\overline{d}$ at low $p_{T}$.

\end{abstract}


\section{Introduction}
In relativistic heavy ion collisions, light nuclei and their
particles can be formed from created nucleons and anti-nucleons or
stopped nucleons ~\cite{coalescence}. Since the binding energy is
small, this formation process can only happen at the late stage of
the evolution when interactions between nucleons and other particles
are weak. This process is called final-state coalescence
~\cite{BA_ref}~\cite{final_state}. Therefore, the production of
light nuclei provides a tool to measure freeze-out properties, such
as particle density ~\cite{density}, correlation volume and
collective motion. Invariant nucleus yield can be related
~\cite{BA_ref} to the primordial yields of nucleons by
Equation~\ref{BA}.
\begin{equation}\label{BA}
    E_{A}{{d^{3}N_{A}}\over{d^{3}p_{A}}} =
B_{A}(E_{p}{{d^{3}N_{p}}\over{d^{3}p_{p}}})^{Z}
(E_{n}{{d^{3}N_{n}}\over{d^{3}p_{n}}})^{A-Z} \approx
B_{A}(E_{p}{{d^{3}N_{p}}\over{d^{3}p_{p}}})^{A}
\end{equation}
where $B_{A}$ is the coalescence parameter.
$E{{d^{3}N}\over{d^{3}p}}$ is the invariant yield of nucleons or
nuclei. A and Z are the atomic mass number and atomic number,
respectively. $p_{A}$ and $p_{p}$ are the momenta of nucleus and
proton where $p_{A} = A\cdot p_{p}$. $B_{A}$ is related to the
freeze-out correlation volume~\cite{BA_ref}: $B_{A}\propto V^{1-A}$.

\section{Experiment and analysis}
\begin{figure}
\begin{center}
\includegraphics[scale=0.48]{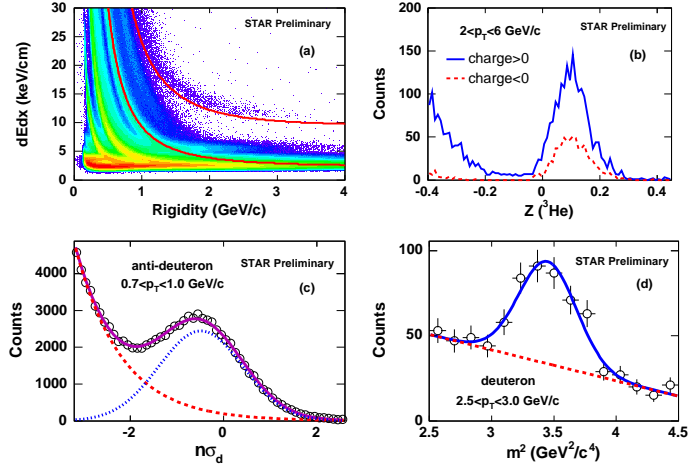}
\caption{(a) TPC $dE/dx$ as a function of rigidity. Lines are
expected values for $d(\overline{d})$ and
$^{3}He(\overline{^{3}He})$ predicted by the Bichsel function. (b) Z
distribution of $^{3}He$ (solid line) and $\overline{^{3}He}$
(dashed line). (c) $n\sigma_{d}$ distribution of $\overline{d}$ at
$0.7<p_{T}<1.0\ GeV/c$ with a Gaussian fit including an exponential
background. (d) $m^{2}$($m^{2}=(p/\beta/\gamma)^{2}$) distribution
for $d$ from TOF after TPC $dE/dx$ selections at $2.5<p_{T}<3.0\
GeV/c$, with a Gaussian fit including a linear background.}
\label{Tech_figure}
\end{center}
\end{figure}
The data presented here are obtained from the Time Projection
Chamber (TPC) and the Time-Of-Flight (TOF) detectors in the STAR
experiment ~\cite{STAR} at RHIC in the year 2004. A data sample of
25 million (16 million for TOF) central triggered events and 24
million (15 million for TOF) minimum-bias triggered events is used
for this analysis. Figure~\ref{Tech_figure} presents the particle
identification techniques and methods. Figure~\ref{Tech_figure} (a)
shows the ionization energy loss ($dE/dx$) of charged tracks as a
function of rigidity ($rigidity=|momentum/charge|$) measured by the
TPC at $-1<\eta<1$. Figure~\ref{Tech_figure} (b) shows $Z$
($Z=\log(dEdx|_{measure}/dEdx|_{predict})$) distribution for
$^{3}He$ and $\overline{^{3}He}$ signals.
After tight track quality selections, the
$^{3}He(\overline{^{3}He})$ signals are essentially background free.
We use counting method to derive the yields.
Figure~\ref{Tech_figure} (c) shows $n\sigma_{d}$ (extracted from
$dE/dx$) distribution for $\overline{d}$ at $0.7<p_{T}<1.0\ GeV/c$.
The signal was fit with a Gaussian function and an exponential
background.
Figure~\ref{Tech_figure} (d) shows $m^{2}$ distribution for $d$ at
$2.5<p_{T}<3.0\ GeV/c$ measured by TOF after the $dE/dx$ selections.
The signal was fit with a Gaussian function and a linear background.
The acceptance and tracking efficiencies were studied by Monto Carlo
GEANT simulations ~\cite{GEANT}. The results presented are corrected
for these effects. Elliptic flow parameter $v_{2}$ was calculated by
the event plane method.

\section{Results}
\begin{figure}
\begin{center}
\includegraphics[scale=0.425]{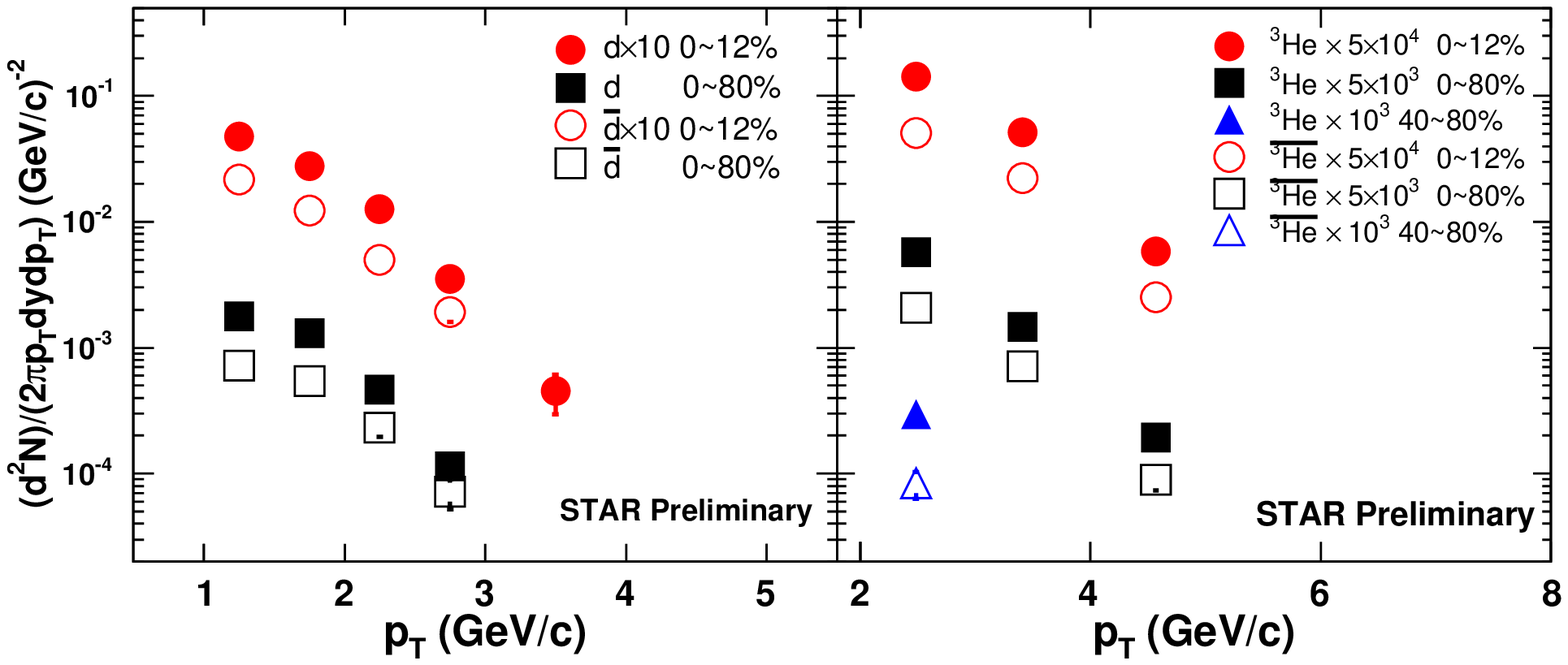}
\includegraphics[scale=0.425]{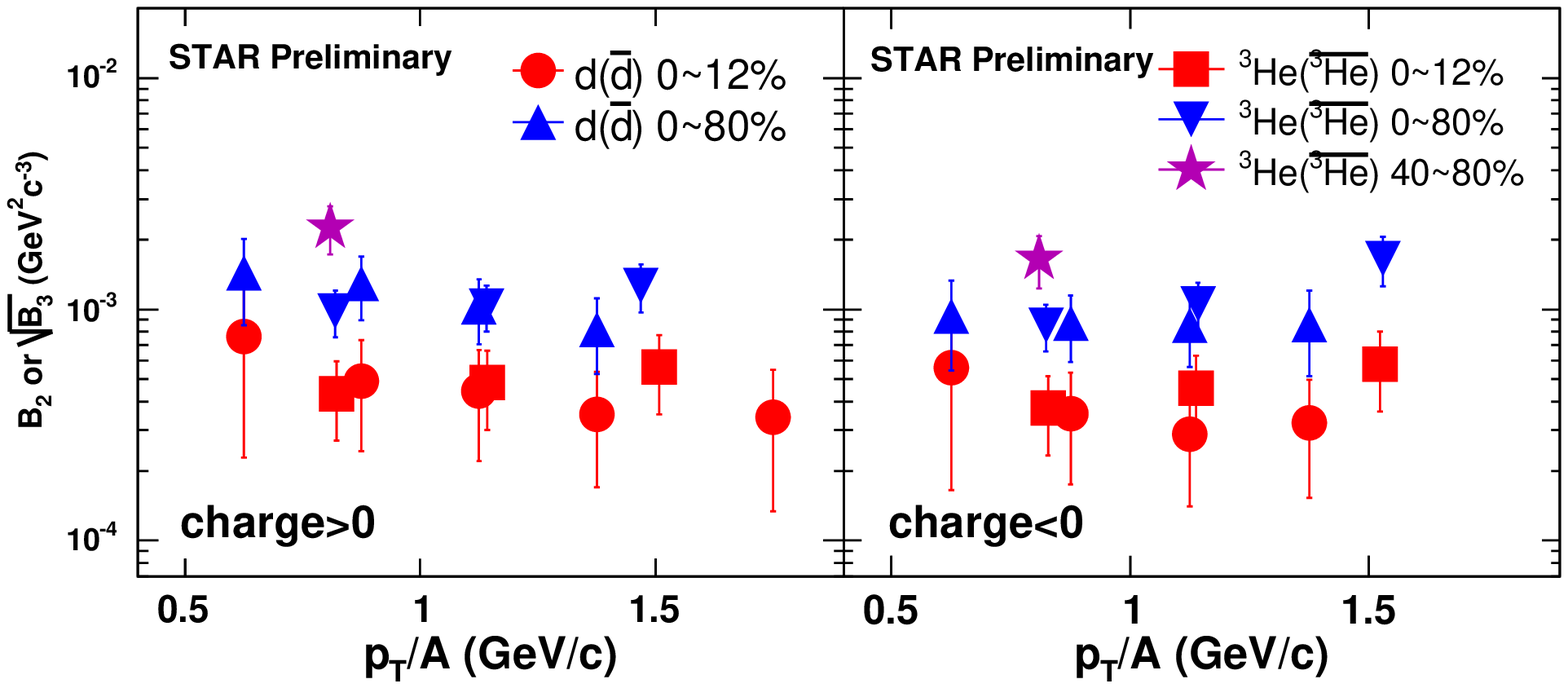}
\caption{\emph{Upper panel}: $p_{T}$ spectra of $d(\overline{d})$
(left panel) and $^{3}He(\overline{^{3}He})$ (right panel) for
different centralities. Solid symbols and open symbols represent the
positive charged particles and negative charged particles
respectively. \emph{Lower panel}: Coalescence parameters $B_{2}$ and
$\sqrt{B_{3}}$ as a function of $p_{T}/A$ for positive charged
particles (left panel) and negative charged particles (right panel).
Errors are statistical only.} \label{spectra_B2B3_figure}
\end{center}
\end{figure}
Figure~\ref{spectra_B2B3_figure} shows the $p_{T}$ spectra and the
extracted coalescence parameters $B_{2}$ and $B_{3}$ for
$d(\overline{d})$ and $^{3}He(\overline{^{3}He})$. Here the proton
and anti-proton spectra are taken from Ref.
~\cite{STAR_proton_spectra}. The $p(\overline{p})$ spectra have been
corrected for feed-down from $\Lambda(\overline{\Lambda}$) and
$\Sigma^{\pm}$ weak decays ~\cite{STAR_proton_spectra}. $B_{2}$ for
$d(\overline{d})$ is consistent with $\sqrt{B_{3}}$ for
$^{3}He(\overline{^{3}He})$, which indicates that the correlation
volumes for $d(\overline{d})$ and $^{3}He(\overline{^{3}He})$ are
similar. Both $B_{2}$ and $B_{3}$ show strong centrality dependence.
In more central collisions, a smaller coalescence parameter
indicates that the correlation volume at thermal freeze-out is
larger and the probability of formation of light nuclei is less.
\begin{figure}
\begin{center}
\includegraphics[scale=0.48]{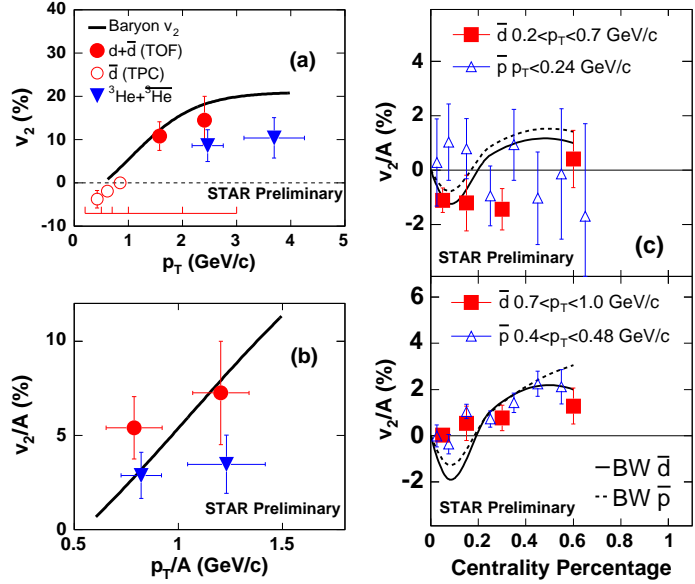}
\caption{(a) The elliptic flow parameter $v_{2}$ from minimum bias
collisions as a function of $p_{T}$ for $^{3}He+\overline{^{3}He}$
(triangles), $d+\overline{d}$ (filled circles) and $\overline{d}$
(open circles); solid line represents the baryon $v_{2}$. (b)
$d+\overline{d}$ and $^{3}He+\overline{^{3}He}$ $v_{2}$ as a
function of $p_{T}$, both $v_{2}$ and $p_{T}$ have been scaled by A.
Errors are statistical only. (c) Low $p_{T}$ $\overline{d}$
$v_{2}/A$ (filled squared) as a function of centrality fraction
($0-10\%,\ 10-20\%,\ 20-40\%,\ 40-80\%$, respectively). Errors are
statistical only. $\overline{p}$ $v_{2}$ is also shown as open
triangles. Blast-wave predictions are show as solids lines
($\overline{d}$) and dashed lines ($\overline{p}$).}
\label{flow_figure}
\end{center}
\end{figure}
Figure~\ref{flow_figure} (a) shows $v_{2}$ as a function of $p_{T}$
for $d+\overline{d}$, $^{3}He+\overline{^{3}He}$ and $\overline{d}$
in minimum-bias collisions. The results with both $v_{2}$ and
$p_{T}$ scaled by A are shown in Figure~\ref{flow_figure} (b). For
comparison, the baryon $v_{2}$ ~\cite{Xin_v2} is also shown as the
solid line. $d+\overline{d}$ and baryon $v_{2}$ seems to follow the
A scaling within errors, indicating that the $d+\overline{d}$ are
formed through the coalescence of $p(\overline{p})$ and
$n(\overline{n})$ just before the thermal freeze-out. The scaled
$^{3}He+\overline{^{3}He}$ $v_{2}$ appears to deviate from the the
baryon $v_{2}$. Further conclusions are limited by poor statistics,
so clearly more data are needed. The $\overline{d}$ $v_{2}/A$ as a
function of centrality fraction is shown in Figure~\ref{flow_figure}
(c). The two panels represent results for two different regions of
$p_{T}$. $\overline{d}$ has a negative $v_{2}$ in central and
mid-central collisions in the transverse momentum range of
$0.2<p_{T}<0.7\ GeV/c$. This negative $v_{2}$ is consistent with a
large radial flow, as the blast-wave predictions show. At the same
$p_{T}/A$ where the $\overline{d}$ is negative, the $\overline{p}$
$v_{2}$ is consistent both with zero and with the $\overline{d}$
$v_{2}$, due to large uncertainties. Though the blast-wave model
predicts the generic feature of negative $v_{2}$, quantitative
agreement between data and model throughout the entire centrality
and $p_{T}$ range is lacking.

\section{Summary}
Taking advantage of the combining STAR TPC and TOF detectors
capabilities, we have measured the $d(\overline{d})$ and
$^{3}He(\overline{^{3}He})$ transverse momentum spectra and elliptic
flow. The value of the coalescence parameters $B_{2}$ and
$\sqrt{B_{3}}$, extracted from the spectral measurements, are found
to be consistent. Systematic studies of $B_{2}$ and $B_{3}$ have
shown decreasing trends as function of collision centrality,
consistent with an increasing source size from peripheral to central
collisions. Comparative analysis of the elliptic flow measurements
show that $d(\overline{d})$ and $p(\overline{p})$ $v_{2}$ scales
with atomic mass number A, which is expected natural consequence of
final-state coalescence. The negative $\overline{d}$ $v_{2}$ values
observed at low $p_{T}$ are consistent with a large radial flow in
the soft sector.

\end{document}